# Effects of alloying elements on carbon diffusion in the austenite (f.c.c.)- and ferrite (b.c.c.)-phases

Zugang Mao,[1] Amir R. Farkoosh[1,2], David N. Seidman[1,2]

[1]Northwestern University, Department of Materials Science and Engineering, Evanston, IL 60208 U.S.A.
[2]Northwestern University Center for Atom-Probe Tomography (NUCAPT), Evanston, IL 60208 U.S.A.

**Abstract**

The effects of alloying elements on diffusion pathways and migration energies of interstitial carbon in austenite (f.c.c.) and ferrite (b.c.c.) are studied using density functional theory first-principles calculations. The binding energies between carbon and alloying elements are determined through 6th nearest-neighbor (NN) distances. The elements studied are Ni, Mo, V, Cr, Mn, Cu, Al, Ti, and Si, relevant to most high-strength steels. Nickel, Mn, Al, and Si have repulsive binding energies; Mo, V, Cr, Cu, and Ti have attractive binding energies in austenite and ferrite. Alloying elements at 1st NN sites of a C atom in an octahedral site introduce asymmetry into the minimum energy diffusion pathway, causing up to ~1 eV changes in saddle-point energies. This pathway goes from one octahedral site to another via intermediate energy states, differing for austenite and ferrite. We find that the elements with attractive binding energies increase the energy barrier for C migration resulting in decelerated carbon diffusion, while the elements with repulsive binding energies decrease the energy barrier for C migration leading to accelerated C diffusion. The magnitude of changes in C migration energies is proportional to the binding energies between C and alloying elements. Among the three austenite stabilizers, Ni and Mn are C diffusion accelerators, while Cu decelerates C diffusion in austenite. Among the four ferrite stabilizers, Si is a C diffusion accelerator, while V and Ti serve as C diffusion decelerators in ferrite. Aluminum has no significant effect on C's diffusivity, while Mo and Cr decelerate C diffusion. (249 words)

## 1. Introduction

In a high-nickel quenched-lamellarized-tempered (QLT) steel with different alloying elements, the nucleation, growth and stability of the austenite (f.c.c.) phase is critical to its strength and toughness. Alloying elements have strong effects on the kinetics and thermodynamics of austenitization in martensitic steels, as they partition between the different phases: ferrite (b.c.c.)/martensite (b.c.t.), austenite (f.c.c.), and carbide precipitates, with different kinetics during the inter-critical QLT treatment [1-3]. The interstitial diffusion of C in different phases [4], which is affected significantly by the presence of substitutional alloying elements [5], often plays a significant role in these thermally activated processes. Understanding C interactions with solute atoms and how these interactions influence C diffusivity in different phases helps to design new heat treatments, which can further improve the mechanical and physical properties of steels. Prior studies [6, 7] suggest that C diffuses by migration from one octahedral site to another nearest-neighbor (NN) octahedral site via a tetrahedral transition site and an intermediate saddle-point site in austenite (f.c.c.). In the relatively simple Fe-C and Fe-C-Mo systems, Wells et al.[8] and Smith [9] measured experimentally C diffusivities in austenite (f.c.c.) and reported a migration energy of ~1.67 eV at 0.1 wt. % carbon. Wert used the classical linear Arrhenius plot to determine the diffusivity of C in ferrite (b.c.c.) between 238 to 473 K and measured a migration energy of ~0.87 eV [10, 11]. Jiang et al. [12] calculated C migration energies of 0.86 and 1.99 eV in b.c.c. and f.c.c. Fe, respectively, using first-principles calculations (0 K). Several prior studies have demonstrated the effects of different alloying elements on C diffusivities, mainly in austenite (f.c.c.), which are contradictory due to the complicated nature of interatomic interactions in multicomponent systems. Krishtal [13] reported that a substitutional alloying element at small concentrations has negligible effects on C diffusion in austenite (f.c.c.). Wada *et al.* [14], Zhukov and Krishtal [15], and Rowan and Sisson [16] found that Ni accelerates and Cr decelerates C's diffusivity in austenite (f.c.c.). Babu and Bhadeshia [17] demonstrated that C's activity can change significantly as a function of substitutional solute atoms, resulting in significant changes in the mobility of C. They established that Ni and Al have a small effect on C's diffusivity in austenite (f.c.c.), whereas Cr and Mo tend to reduce C's diffusivity. By extending the Siller and McLellan model [18], used for austenite (f.c.c.) containing substitutional solute atoms, as well as C, Babu and Bhadeshia [17] developed a model capable of predicting C's diffusivity in ternary Fe-C-X (X = substitutional element) systems. This model was, in general, in a reasonable agreement with experiments. A large discrepancy between experiments and predictions was, however, found in the Fe-C-Cr system. The reason for this discrepancy is unclear. Utilizing experimental diffusion data and regression analyses, Lee *et al.* [2] proposed an entirely empirical model to predict C's diffusivity in austenite (f.c.c.) as a function of chemical composition and temperature for Fe-C based multicomponent steels. Their results indicate that austenite-stabilizing elements enhance C's diffusivity in austenite (f.c.c.), while carbide forming elements decrease C's diffusivity.

In this contribution, we perform first-principles calculations, within the framework of density functional theory (DFT), to gain a fundamental understanding of C diffusion processes in steels containing different substitutional alloying elements. We begin with a DFT model which is capable of determining C diffusion pathways in f.c.c. and b.c.c iron and then consider extensions to this model to investigate the effects of the alloying elements: Ni, Mo, V, Cr, Mn, Cu, Al, Ti, and Si on C's migration energies in f.c.c. and b.c.c. iron, which is relevant to a 10 wt.% Ni steel developed for Naval applications [3]. These substitutional elements are divided into three groups: (i) austenite stabilizers, which partition to the austenite (f.c.c.) phase - Ni, Mn, and Cu; (ii) ferrite stabilizers, which partition to the ferrite (b.c.c.) phase - Si, V, Al, Ti; and (iii) termed hereafter as "vacillating elements" with half-filled (five unpaired) d-electrons [(n-1) $d^5ns^1$], which partition to either austenite (f.c.c.) or ferrite (b.c.c.), depending on their concentrations and temperature - Cr and Mo. Having different binding energies with C atoms, substitutional alloying elements can alter the migration energies of C atoms. Herein, we first study the partitioning behavior of these elements experimentally with atom-probe tomography. Then we calculate the binding energies between the alloying elements and C atoms through $6^{th}$ nearest-neighbor (NN) distances in ferrite (b.c.c.) and austenite (f.c.c.). Finally, the minimum-energy diffusion pathways for the migration of a diffusing C atom are determined and the corresponding migration energy barriers for C diffusion are calculated considering the effects of alloying elements for all the possible $1^{st}$ NN sites.

## 2. Experimental Procedure and Theoretical method

### 2.1. Steel preparation and atom-probe tomography analyses

The chemical composition of the 10 wt.% Ni steel, determined by optical emission spectroscopy and combustion infrared detection analyzer (for the metallic elements and carbon, respectively), is presented in Table 1. This steel was prepared in a vacuum induction furnace and hot rolled at ~845 °C after homogenization at 1260 °C for 7 h under a protective nitrogen gas. The Quenching-Lamellarization-Tempering (QLT) heat-treatment was employed, which consists of an austenitizing treatment at 800 °C for ~1 h (Q), followed by the first intercritical heat treatment (L) at 650 °C for ~ 30 minutes, and a second intercritical heat treatment (T) at 590 °C for 1 h. Each heating step is terminated by quenching the specimen to room temperature. This heat treatment creates a significant amount ($V_f$ ~18%) of thermally and mechanically stable retained austenite (f.c.c.) grains (D ~50-100 nm) within a tempered martensitic matrix [3]. Nanotips for three-dimensional (3D) local-electrode atom-probe (LEAP) tomography investigations were prepared by a site-specific focused-ion beam (FIB) microscopy lift-out technique. The 3-D LEAP tomographic experiments were performed utilizing a laser-pulsed LEAP5000 XS tomograph (Cameca Instruments Inc., Madison, WI) at 60 K in ultrahigh vacuum (<$10^{-8}$ Pa). Picosecond ultraviolet (UV) laser pulses (wavelength = 355 nm) were applied with an energy of 30 pJ per pulse and a pulse repetition rate of 500 kHz, while maintaining an average detection rate (number of ions per pulse) of 2%. Data analyses were performed using the program IVAS 3.8 (Cameca, Madison, WI) and the proximity histogram methodology [19].

**Table 1**. Chemical composition of the 10 wt.% Ni steel

| 10 % Ni steel | C | Ni | Mn | Si | Cr | Mo | Cu | V | Al |
|---|---|---|---|---|---|---|---|---|---|
| Wt. % | 0.1 | 10 | 0.60 | 0.2 | 0.6 | 1.23 | 0.15 | 0.08 | 0.03 |
| At. % | 0.47 | 9.55 | 0.61 | 0.40 | 0.65 | 0.72 | 0.13 | 0.09 | 0.06 |

## 2.2. Theoretical method

First-principles calculation are performed utilizing VASP (Vienna Ab initio Simulation Program), which employs density-functional theory (DFT) within the projector-augmented wave (PAW) method [20-24]. The generalized gradient approximation (GGA) formulated by Perdew, Burke, and Ernzerh (PBE) [25] is employed for the exchange and correlation energy terms, as GGA describes spin-polarized Fe better than the local-density approximation (LDA). The Vosko-Wilk-Nusair method [26] is employed in the spin interpolation of the correlation energy. The cutoff energy of the augmentation functions is 700 eV. The electronic wave functions are sampled on 16×16×16 k-points. The local magnetic moments on atoms are initialized to impose magnetic state ordering and are then allowed to relax. The relaxed local magnetic moments are determined by integrating the spin density within spheres centered on atoms. A 3x3x3 supercell is utilized for the f.c.c. structure and a 4x4x4 supercell is employed for the b.c.c. structure for binding energy calculations. The equilibrium supercell volumes at 0 K and ground-state energies are obtained utilizing Murnaghan's equation of state [27] and a set of energy vs. volume data. Due to the four unpaired valence electrons in the active electron orbital of the outer shell, Fe can demonstrate different magnetic phenomena in its different forms, such as atomic clusters, surfaces, thin films, and different bulk structures with no vacancies. The prior first-principles studies successfully describe the magnetic phases of bulk and thin-film Fe. In austenite (f.c.c.), three possible magnetic states are investigated: the single-layer antiferromagnetic state (AFMS), the face-centered tetragonal (f.c.t.) state, the double-layer antiferromagnetic (AFMD) f.c.t. state, and f.c.t. ferromagnetic high-spin (FMHS) calculations. The antiferromagnetic single-layer (AFMS) phase refers to the common antiferromagnetic structure (↑↓↑↓), and the antiferromagnetic double-layer (AFMD) refers to the double-layer antiferromagnetic structure (↑↑↓↓) [28]. Ferromagnetic high-spin (FMHS) refers to a relatively higher-energy high-spin ferromagnetic structure (↑↑↑↑). Experimentally, f.c.c. iron is found to be paramagnetic in the temperature range 900-1400 K. Because the paramagnetic state consists of randomly disordered local magnetic moments in austenite (f.c.c.), randomly distributed ferromagnetic-domains can represent accurately the paramagnetic state: we chose the most possible ferromagnetic high-spin (FMHS) state for our calculations [28, 29]. The optimized lattice constant and local magnetic moment of the ferromagnetic FMHS austenite (f.c.c.), are 3.640 Å and 2.44 $\mu_B$/atom, respectively, agreeing with an experimental value of 3.65 Å for the lattice constant. The lattice constant and local magnetic moment of ferromagnetic ferrite (f.c.c.) are 2.834 Å and 2.15 $\mu_B$/atom, agreeing with the values 2.86 Å for the lattice constant at room temperature and 2.22 $\mu_B$/atom for the magnetic moment [30].

The nudged elastic band (NEB) method [31] is employed utilizing VASP to determine the minimum-energy paths for C solute atoms transitioning between the octahedral interstitial sites, by calculating the associated energy barriers in bulk f.c.c. and b.c.c. Fe. In this method, an interpolated series of configurations between the initial and final states (so-called "images") are connected by "springs" and are simultaneously fully relaxed. With appropriate 2D projections, the true force and the spring force acting on each image are separated from the total force to determine energy minimization path. This procedure determines the migration path and leads to the migration energy computed as the minimum energy along this path. For C migration in austenite (f.c.c.), the initial projection is a set of non-parallel linear interpolation images between the initial and final states with 19 configurational images for a $Fe_{64}C_1$ supercell. The initial projection converges to the minimum-energy path within 200 ionic iterations. For C migration in ferrite (b.c.c.), a $Fe_{128}C$ supercell is used for the minimum-energy path calculation. A set of linear interpolation images between the initial and final states with 13 images converges to a minimum-energy path within 200 ionic iterations.

## 3. Results and discussion

### 3.1. The experimental concentration profiles and partitioning behavior of alloying elements

The concentrations and partitioning behaviors of the alloying elements determine the NN elemental environments around a C atom, which critically affect its diffusivity. Solute atoms partition to austenite (f.c.c.) and ferrite (b.c.c.) or tempered martensite with different tendencies depending on their electronic structure, concentration, and their interatomic interactions with the other solute atoms [32]. In general, common austenite stabilizers, such as Ni, Mn, Cu and C partition to the austenite (f.c.c.) phase, and the elements, such as Si, V, Ti, Mo and Cr partition to the ferrite (b.c.c.) phase. This behavior changes, however, in the presence of certain elements, specifically for Cr and Mo (vacillating elements). In a super duplex stainless steel, Weber et al. [33] observed that Mo partitions significantly to the ferrite (b.c.c.) phase, while the partitioning of Cr to ferrite (b.c.c.) is small. The presence of Ni increases the partitioning level of Mo and Cr, while in the presence of nitrogen partitioning of these elements diminishes. Cortie and Potgieter [34] have reported a similar behavior in a stainless steel. In the binary Al-Cr system, the partitioning behavior of Cr is concentration dependent [35]. Using atom-probe tomography (APT), Xiong et al. [36] have demonstrated that in a low C steel with 1.6 wt.% Mn, Cr partitions to the retained austenite (f.c.c.) along with Mn.

Figure 1(a) displays the 3-D APT reconstruction of a quenching-lamellarization-tempering (QLT)-treated sample of a 10 wt.% Ni steel, Table 1, which contains two phases with dramatically different Ni concentrations separated by 12 at.% Ni isoconcentration surfaces, as well as $M_2C$ (M = Mo, Cr, V) carbide precipitates. Our prior investigations [37-40] demonstrated that the Ni-rich phase, colored green, is austenite (f.c.c.), which forms at martensite lath boundaries. The partitioning of the alloying elements is quantitatively shown in Fig. 1b using proximity histograms (or proxigrams for short), which calculates the concentration of each element as a function of distance from a specified (12 at.% Ni) isoconcentration surface. Metal carbides (shown in red) are excluded from the calculations. Nickel, Mn, Cu, C and Cr partitioning to the austenite (f.c.c.) phase, and V partitioning to the martensite (b.c.t) phase are observed, Fig. 1b. There isn't a significant partitioning preference for Si. A significant segregation of C and Mo at the heterophase interface between martensite (b.c.t.) and austenite (f.c.c.) are observed.

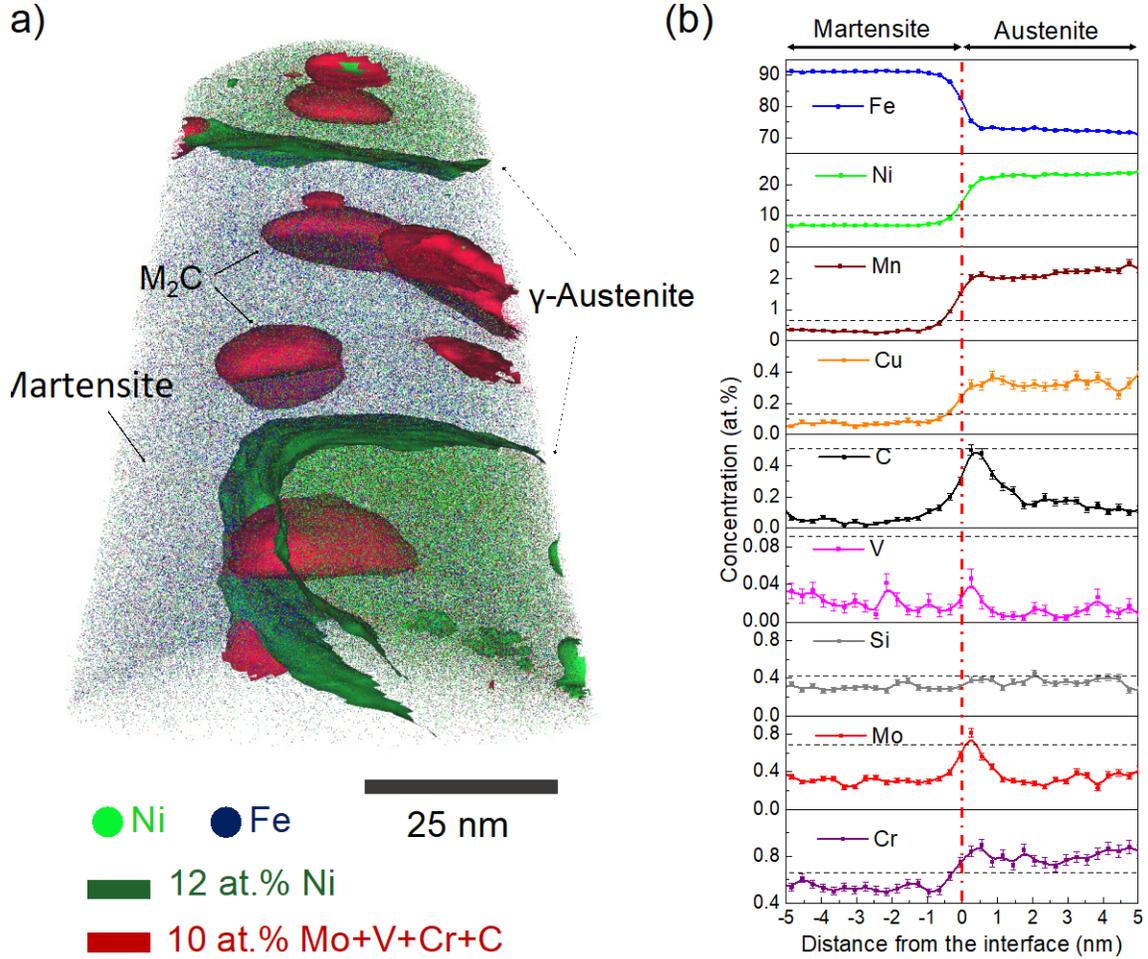

**Fig. 1** (a) 3-D APT reconstruction of a QLT-treated 10 wt.% Ni steel. The Ni-rich regions (γ-Austenite (f.c.c.)) are delineated by a 12 at.% Ni isoconcentration surface. Only Fe and Ni atoms are displayed for the purpose of clarity. Metal carbides are delineated by 10 at.% (Mo plus V plus Cr plus C) isoconcentration surfaces. (b) Proximity histogram concentration profiles obtained from the isoconcentration surfaces in (a) displaying the concentration profiles of Fe, Ni, Mn, Cu, C, V, Si Mo and Cr across the austenite-martensite heterophase interface. The elements Ni, Cu, C, Cr all partition to the austenite (f.c.c.) phase.

### 3.2. The binding energy conditions between alloying elements and carbon atoms

We first examine the interatomic binding conditions of the alloying elements with C (C-M pairs), which have major effects on C diffusion in the ferrite (b.c.c.) and austenite (f.c.c.) phases. The binding energies of alloying elements (M) with carbon for an i$^{th}$ NN distance are calculated in the austenite (f.c.c.) and ferrite (b.c.c.) matrices employing the following equations:

$$E_{bcc, C-M}^{b, i} = E_{C+M}^{bcc, i} - \left( E_C^{bcc} + E_{M \to Fe}^{bcc} \right) \qquad (1)$$

$$E_{fcc,\,C-M}^{b,\,i} = E_{C+M}^{fcc,\,i} - \left(E_C^{fcc} + E_{M\to Fe}^{fcc}\right) \tag{2}$$

Where $E_{C+M}^{bcc,\,i}$ and $E_{C+M}^{fcc,\,i}$ are the total energies with a C atom and an alloying element at an $i^{th}$ NN distance in b.c.c. and f.c.c. Fe phases, respectively; $E_C^{bcc}$ and $E_C^{fcc}$ are the total energies with a C atom in an octahedral interstitial site in b.c.c. and f.c.c. Fe, respectively; and $E_{M\to Fe}^{bcc}$ and $E_{M\to Fe}^{fcc}$ are the total energies with the alloying elements in b.c.c. and f.c.c. Fe phases, respectively.

The calculated binding energies of the alloying elements with C in an austenite matrix (FMHS state) are displayed in Fig. 2. *In all our calculations a negative binding energy is attractive, and a positive binding energy is repulsive.* In the austenite (f.c.c.) phase, in the 1$^{st}$ NN position, we discovered that among the austenite stabilizing elements, Ni and Mn, have positive (repulsive) binding energies, while Cu has a negative (attractive) binding energy. While the ferrite stabilizing elements, V and Ti, as well as vacillating elements Mo and Cr have negative (attractive) binding energies, whereas Si and Al have positive (repulsive) binding energies. In the ferrite (b.c.c.) phase, the trends for the binding energies of alloying elements with C are similar to those in austenite (f.c.c.). The details are displayed in Fig. 3. We find that the values of the binding energies in ferrite (b.c.c.) are, in general, smaller than those in austenite (f.c.c.), which is due to more free volume. The C-Ni, C-Mn, C-Si, and C-Al atomic pairs have positive (repulsive) binding energies, whereas C-V, C-Ti, C-Cu, C-Mo, and C-Cr have negative (attractive) binding energies in the 1$^{st}$ NN position.

These results indicate that the strongest repulsive binding energy (0.45 eV/pair) is for the C-Si pair, while the strongest attractive binding energy (0.45 eV/pair) is for the C-Ti pair in the austenite (f.c.c.) phase. We also find that certain carbide-forming elements, that is Cr and Mo, have weak attractive binding energies with C in the 1$^{st}$ NN position comparable, for instance, to Cu which does not form carbide precipitates in steels. The oscillating nature of the M-C binding energies observed in Figs. 2-3 are consistent with theoretical models for solute atom interactions in metallic systems [41, 42]. The amplitudes of these Friedel-like oscillations decay significantly with increasing the NN distance. This behavior has been observed in different metallic systems including Al-transition-metals [43-45] and Fe-transition-metals [46]. All the calculated binding energies are at 0 K, which implies that there is not an entropic term, i.e., -T$\Delta$S, where $\Delta$S is the entropy change, which is temperature dependent.

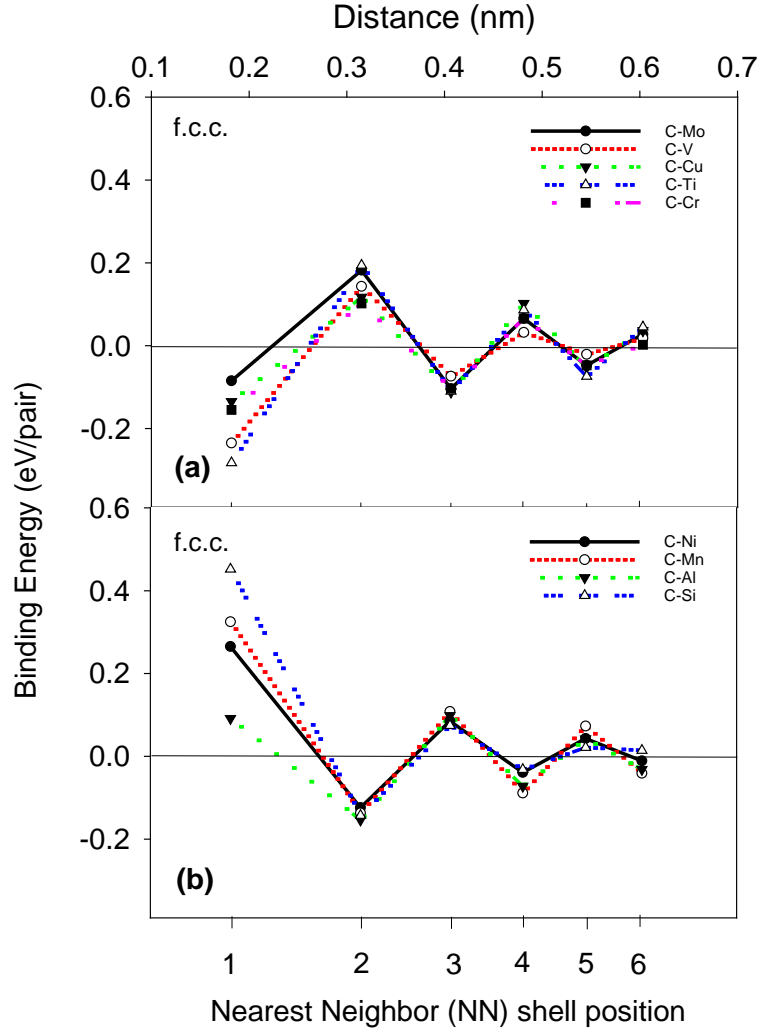

**Fig. 2** First-principles calculated atomic binding energies with carbon for different alloying elements at different nearest-neighbor (NN) distances in ferromagnetic high-spin (FMHS) austenite (f.c.c.), unit: eV/pair: (a) elements with an attractive binding energy with C in 1st NN positions; (b) elements with a repulsive binding energy with carbon in 1st NN positions. The binding energies all decrease with increasing NN distance.

### 3.3. The migration paths of carbon atoms

Now, we discuss the diffusion mechanism for C atoms in the austenite (f.c.c.) and ferrite (b.c.c.) phases. We study different transition states of C migration via interstitial sites with and without the alloying elements located at the 1st NN distance of the C atoms. We ignore the C-solute interactions beyond the 1st NN distance as they are relatively weak as demonstrated in Figs. 2-3. These weak interactions are frequently ignored in standard diffusion models [47].

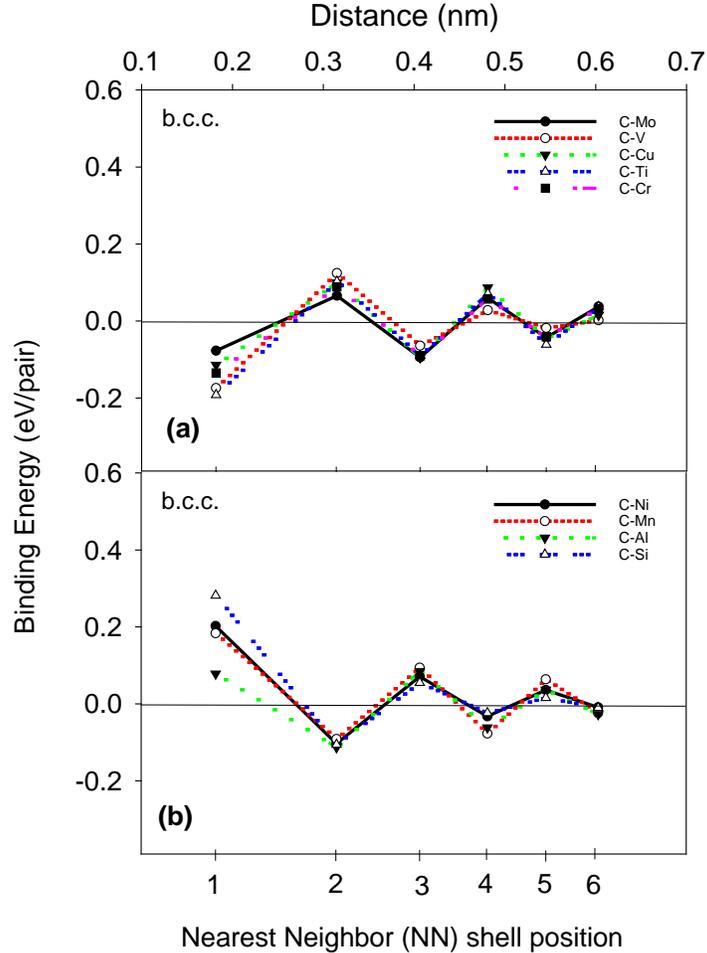

**Fig. 3** First-principles calculated atomic binding energies with C, for different alloying elements at different nearest-neighbor (NN) distances in ferrite (b.c.c.), unit: eV/pair. (a) elements with an attractive binding energy with carbon in 1st NN positions; (b) elements with a repulsive binding energy with C in 1st NN positions.

First, we demonstrate that a possible migration pathway is a minimum-energy pathway. Figure 4 displays the five states for migration of a C atom (solid-black circle surrounded by an annular red ring) in austenite (f.c.c.) from an octahedral site (an edge center of the unit cell) to a NN octahedral site (body-center of the unit cell) via a tetrahedral site. They are calculated employing the NEB methodology using the following scheme: (i) the initial state, Fig. 4a; (ii) the first intermediate saddle-point state, approximately in the center of a plane going through Fe1, Fe2, and Fe3, Fig. 4b; (iii) a local energy minimum at a tetrahedral interstitial site, consisting of Fe1, Fe2, Fe3, and Fe4, Fig. 4b; (iv) a second intermediate saddle-point state approximately in the center of a plane going through Fe2, Fe3, and Fe4, Fig. 4b; and (v) the final state, Fig. 4c. The NEB calculations, Fig. 5, demonstrate that, in the absence of substitutional alloying elements, the C diffusion path follows a symmetric pattern with two maximum saddle points, 1.82 eV, and one local minimum of 1.53 eV in between the two saddle points at a tetrahedral transition site.

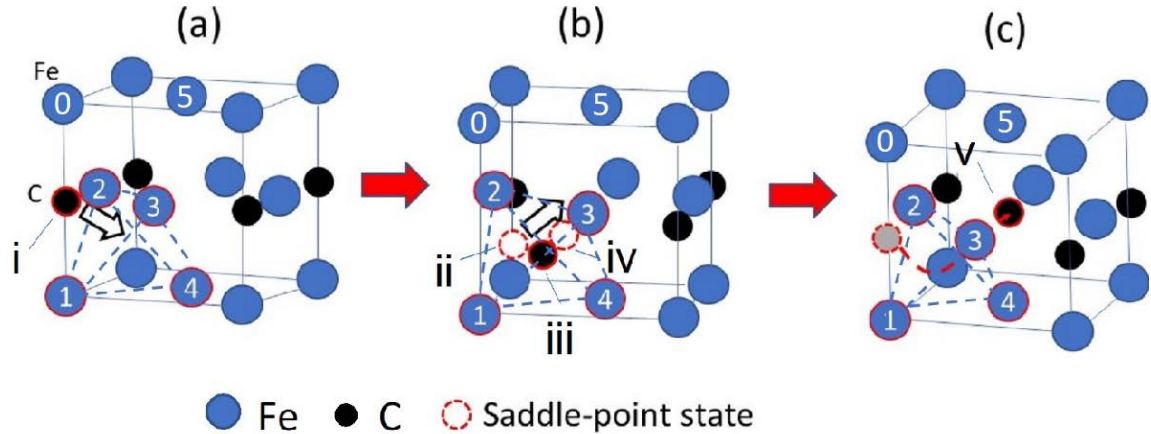

**Fig. 4** The diffusion pathway of a C atom determined with NEB method and four possible 1st NN alloying effective sites in austenite (f.c.c.): (a) The initial state (i) of a migrating black C atom, which is encircled in red; (b) The first intermediate (ii), between Fe1, Fe2 and Fe3, saddle-point (iii), tetrahedral site between Fe1, Fe2, Fe3 and Fe4 and the second intermediate (iv), between Fe2, Fe3 and Fe4 sates; and (c) The final state (v), which is identical to the initial state.

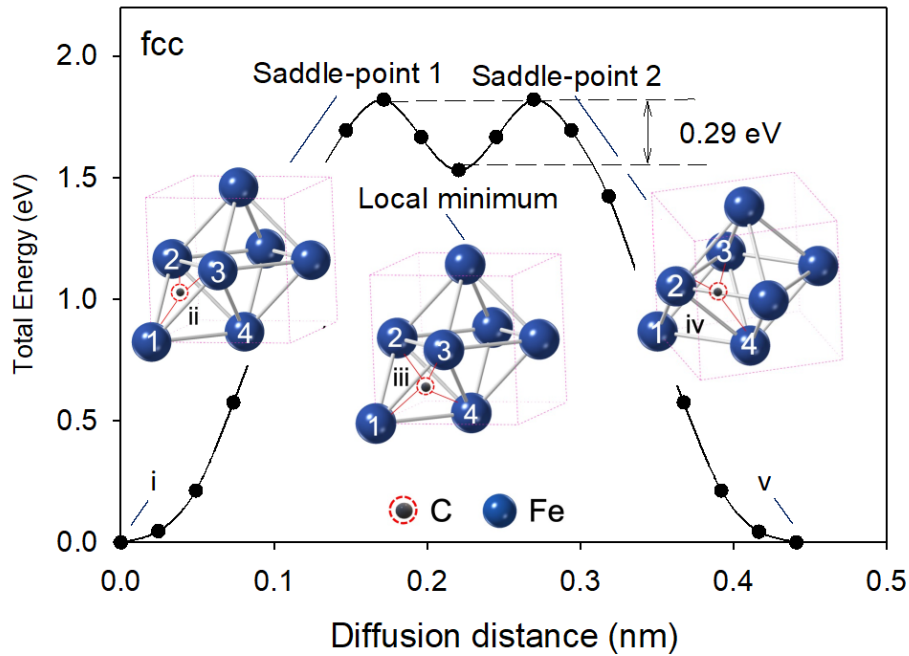

**Fig. 5** The migration energy of a diffusing carbon atom following the minimum-energy path for carbon diffusion in austenite (f.c.c.). It includes five states: the initial state (i), the first saddle-point state, 1.82 eV (ii), the intermediate local minimum energy state, 1.53 eV (iii), the second saddle-point state, 1.82 eV (iv), and the final state (v). The intermediate local minimum energy state is 0.29 eV lower than the two saddle-point configurations.

Similar calculations for a C atom migrating in ferrite (b.c.c.) are performed utilizing the five migration states displayed in Fig. 6. Due to the tetragonality of the relaxed supercells (with C occupying the octahedral site) and to ensure that the initial and final states have the same cell shape and volume, the C atom migration path is as follows [48]: (i) the initial state, C atom in an octahedral site, Fig. 6a; (ii) the first saddle-point state, C atom in a tetrahedral site, Fig. 6b; (iii) the local minimum, C atom in the 1$^{st}$ NN octahedral transition site, Fig. 6a; (iv) the second intermediate saddle-point state, C atom in a tetrahedral site, Fig. 6b; and (v) the final state, C atom in an octahedral site, Fig. 6c. This model also permits us to study the effects of alloying elements when they substitute one of the 6 possible NN lattice sites around the C atom, on C migration energies in ferrite (f.c.c.). The corresponding migration energies of a C atom diffusing in ferrite (b.c.c), without alloying elements are displayed in Fig. 7. In our DFT model, the C atom's migration path follows a symmetric pattern with two saddle points, 0.98 eV, and an intermediate local energy minimum state, at almost the same energy level as the initial and final states. The C atom's migration path in ferrite (b.c.c.) is different from its path in austenite (f.c.c.). The C atom migrates from a distorted octahedral site, via a distorted tetrahedral site (the saddle-point), to the NN distorted octahedral site (between Fe4 and Fe6), Fig. 6(b). The C atom then moves into a second saddle-point state (a distorted tetrahedral site) before entering the final distorted octahedral site, Fig. 6(c). In the absence of alloying elements, the calculated migration energy of a C atom is 1.82 eV/atom in austenite (f.c.c.) and 0.98 eV/atom in ferrite (b.c.c.), which indicates that C atoms migrate more readily in ferrite (b.c.c.) than in austenite (f.c.c.) due to its smaller migration energy as is well known.

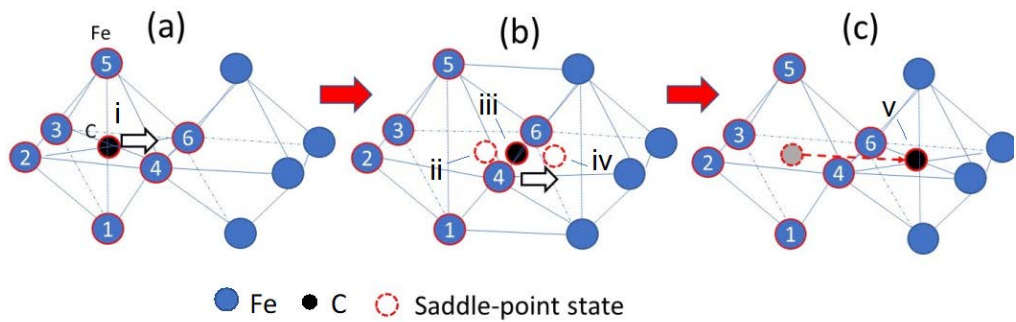

**Fig. 6** The pathway for diffusion of C in ferrite (b.c.c.), and six possible 1$^{st}$ NN lattice sites around a C atom in an octahedral interstitial site: (a) The initial state of a carbon atom (i); (b) The first saddle-point state (ii), the intermediate local minimum state (iii) and the second saddle-point state (iv); and (c) The final state (v). The diffusing C atom is in solid black, and it is encircled in red.

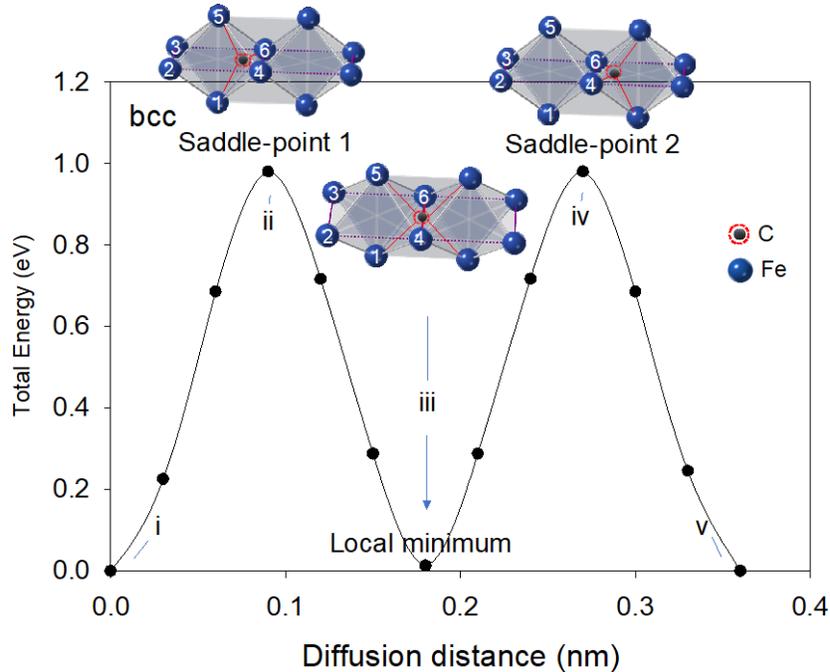

**Fig. 7** The migration energy of a diffusing C atom following the minimum-energy path for C diffusion in ferrite (b.c.c.). It includes five states: the initial state (i), the first saddle-point state (ii), the intermediate local minimum state (iii), the second saddle-point state (iv), and the final state (v). The intermediate local energy minimum state is at the same energy level as the initial- and final-stages.

### 3.4. The effects of alloying elements on C's diffusion

Beyond the $1^{st}$ NN distance, C atom's interactions are relatively weak, Figs. 2-3, and are therefore ignored in our migration energy calculations. On the basis of their interaction energies with C solute atoms at the $1^{st}$ NN distance, Figs. 2-3, we divide all the alloying elements studied into two main groups: (1) those with an attractive binding energy with C; and (2) those with a repulsive binding energy with C. All the possible $1^{st}$ NN lattice sites in the vicinity of a C along the diffusion path are considered for both austenite (f.c.c.) and ferrite (b.c.c.). The same migration paths are assumed for a carbon atom as in the binary Fe-C system, Figs. 5&7, with, however, different *energy barriers* when the substitutional alloying elements occupy one of the $1^{st}$ NN sites. We first discuss both cases in austenite (f.c.c.).

### 3.4.1. Carbon diffusion in f.c.c. austenite under the effects of alloying elements

In austenite (f.c.c.), five main possibilities arise from the substitution of one of the NN sites along the carbon migration path with a substitutional solute atom. We have categorized these five distinct types of 1st NN lattice sites based on their interactions with the carbon atom as it migrates from an initial octahedral site to the NN octahedral site via an intermediate tetrahedral site, Fig. 4.

*Type 1 sites (e.g. Fe0):* as a carbon atom enters the intermediate tetrahedral site via the Fe1-Fe2-Fe3 plane, it "breaks bonds" with atoms residing in *Type 1* sites.

*Type 2 sites, (e.g Fe1 on the Fe1-Fe2-Fe3 plane):* the migrating C atom breaks bonds with atoms residing in *Type 2* sites upon exiting the tetrahedral site.

*Type 3 sites (e.g. F2 or F3, on both Fe1-Fe2-Fe3 and F2-F3-F4 planes)*: atoms residing in these sites do not break bonds with the migrating carbon atom.

*Type 4 sites, (e.g. Fe4, on the F2-F3-F4 plane)*: as the carbon atom enters the tetrahedral site, it "makes new bonds" with the atoms residing in *Type 4* sites.

*Type 5 sites, (e.g. Fe5)*: existing the tetrahedral site, the migrating carbon atom enters the NN octahedral site, forming new bonds with atoms residing in the *Type 5* site.

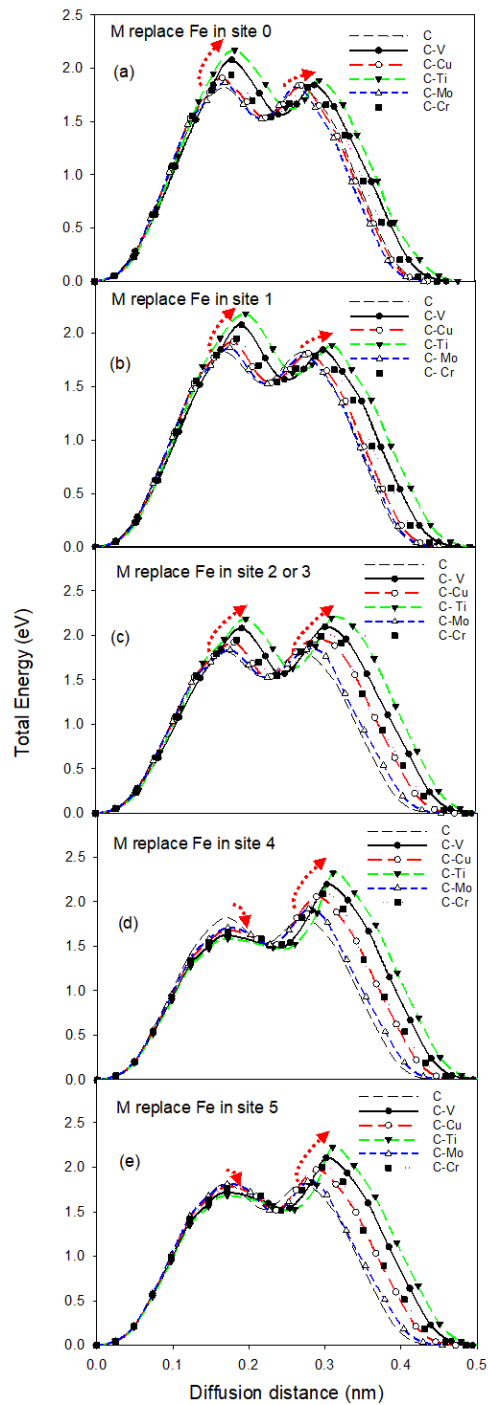

**Fig. 8** The effects of alloying elements with attractive C-M (V, Cu, Ti Mo, Cr) binding energies on the migration energies of a carbon atom in austenite (f.c.c.): An alloying element substituting on (a) site Fe0; (b) site Fe1; (c) sites Fe2 or Fe3; (d) site Fe4; and (e) site Fe5. The presence of substitutional solute atoms, M, in the 1$^{st}$ NN distance of the C atom introduces an asymmetry into the total energy versus diffusion distance plot, the magnitude of which is a function of C-M binding energy given in Fig. 2. The red arrows indicate increasing or decreasing trends in the saddle-point energies.

*Effects of substitutional elements (V, Cu, Ti, Mo, Cr) with attractive C-M binding energies:* We replace one Fe atom in each of the *Type1-5* sites (i.e., Fe0, Fe1, Fe2 or Fe3, Fe4 and Fe5) with one solute atom at a time and then re-calculate the total energy for the minimum-energy paths as displayed in Fig. 8. When an alloying element with an attractive C-M binding energy replaces Fe0 or Fe1, Fig. 8a-b, we find that the energies of the saddle-points increase significantly compared with those in binary Fe-C system (1.82 eV). This increase in the saddle-point energy, which is more significant for the first saddle-point than the second one, scales with the C-M binding energy, Table 2. For instance, the C-Ti pair in austenite (f.c.c.) has the largest attractive binding energy at the 1$^{st}$ NN site, 0.28 eV, and correspondingly the first saddle-point energy is 2.19 eV, which is 0.37 eV greater than in the binary Fe-C system. The C-Mo pair, however, has a weak attractive binding energy of 0.09 eV and the corresponding saddle-point energy is only 1.86 eV. We note that the C migration distance for the minimum-energy path increases slightly (~ 0.03-0.04 nm) when alloying elements are present at sites Fe0 and Fe1.

When an alloying element with an attractive C-M binding energy replaces Fe2 or Fe3, Fig. 8c, both saddle-point energies increase above 1.82 eV with similar trends, which also scale with the C-M binding energies The C migration distance for the minimum-energy path increases slightly (~ 0.04-0.05 nm) with an alloying element is present at sites Fe2 or Fe3.

In the case of the Fe 4 and Fe 5 sites, Fig. 8 d-e, the second saddle-point energy increases significantly in the presence of alloying elements, whereas the first saddle-point energy decreases slightly, resulting in a significantly asymmetric migration path. In this case, the changes in the saddle-point energies are also proportional to the C-M binding energies and the C migration distance for the minimum-energy path increases slightly (~ 0.04 nm) with alloying elements present at sites Fe4 and Fe5.

These results indicate that in the presence of the alloying elements with attractive C-M (Ti, Cu, V, Cr, and Mo) binding energies a higher energy cost is required for the C atom to enter (higher first saddle-point energies, Figs. 8a-c) or exit (higher second saddle-point energies, Figs. 8-d-c) the intermediate tetrahedral site along its migration path. Thus, we conclude that these elements decrease C diffusion in the austenite (f.c.c.) phase due to a higher migration energy for diffusion, which agrees well with the experimental data [13-16], Table 2.

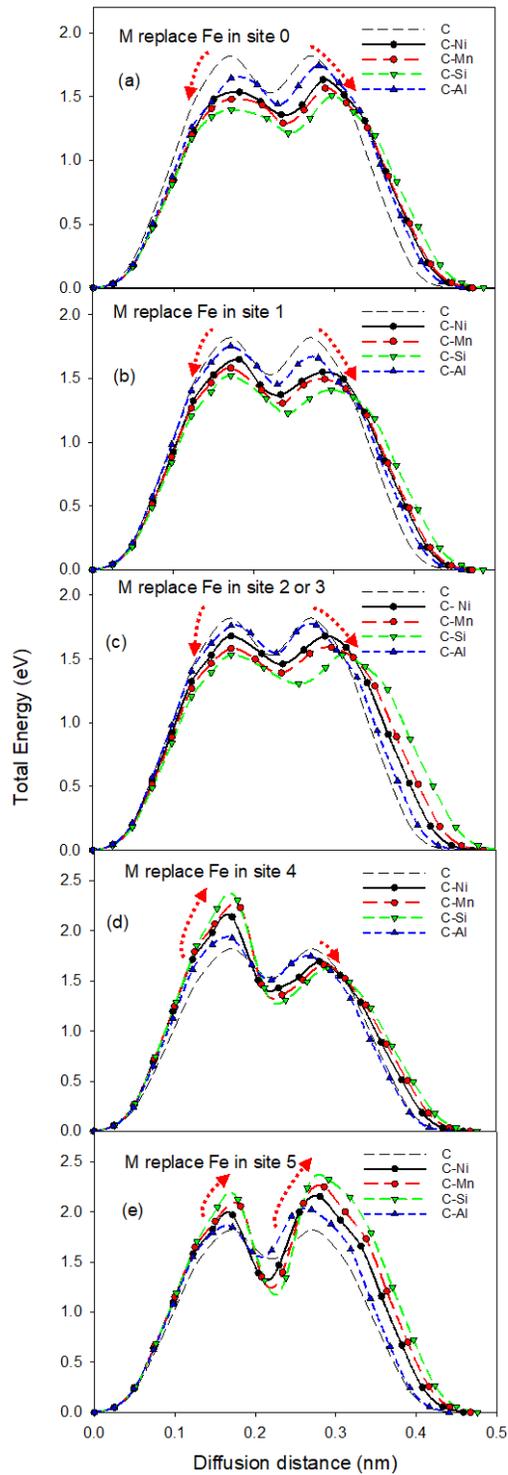

**Fig. 9** The effects of alloying elements with repulsive C-M (Ni, Mn, Si, Al) binding energies on the migration energies of a carbon atom in austenite (f.c.c.): An alloying element substituting on (a) site Fe0; (b) site Fe1; (c) sites Fe2 or Fe3; (d) site Fe4; and (e) site Fe5. The C-M binding energies introduce an asymmetry into the total energy versus diffusion distance plot. The red arrows indicate increasing or decreasing trends in the saddle-point energies.

*Effects of substitutional elements (Ni, Mn, Si, Al) with repulsive C-M binding energies:* For the elements with repulsive C-M binding energies (M = Ni, Mn, Si, Al), the minimum-energy path for C migration in austenite (f.c.c.) is displayed in Fig. 9. A repulsive C-M binding energy decreases the probability of a substitutional atom, M, being at the $1^{st}$ NN distance of the C atoms in dilute alloys. Thus, it is anticipated that the influence of these elements on the diffusivity of C is minimal. In steels with relatively high concentrations of solute atoms (e.g., austenite phase in Fig. 1), these elements, however, can significantly affect the diffusivity of C atoms as demonstrated, for example, for Si [49].

Our calculations reveal that in this case, the migration energy barriers are completely different between the cases when the solute atoms occupy different types of the NN lattice sites along the migration path. In the case of solute atoms replacing one of the Fe atoms in the Fe0, Fe1, Fe2 or Fe3 sites, Figs. 9a-c, the saddle-point energies decrease significantly compared with the base case without an alloying element. The decrease is strongly affected by the C-M binding energy. The larger the repulsive binding energy the alloying element has with a C atom the smaller is the corresponding saddle-point energy. For example, the C-Si pair has the largest repulsive binding energy (0.45 eV), while the second saddle-point energy is 1.41 eV in Fig. 9b. On the other hand, the C-Al pair has a small repulsive binding energy, 0.09 eV, while the second saddle-point energy is 1.66 eV, Fig. 9b.

In case of the Fe4 sites, Fig. 9d, we find that the first saddle-point energy increases significantly in the presence of alloying elements (from 1.82 eV to 2.31 and 1.92 eV for M = Si and Al, respectively), while the second saddle-point energy decreases slightly (from 1.82 eV to 1.22 and 1.75 eV for M = Si and Al, respectively).

In case of the Fe5 sites, Fig. 9e, both saddle-point energies increase in the presence of the alloying elements. The intermediate local minimum energy, however, decreases which may result in a longer residency time of the C atom in the local minimum energy state in the intermediate tetrahedral site.

These results indicate that in austenite (f.c.c.) the easy (minimum) energy path for carbon's migration is via circumventing the intermediate tetrahedral sites (Fe1-Fe2-Fe3-Fe4, Fig. 4a) with an alloying element present in the *Type4* (Fe4) sites and the second octahedral sites with alloying elements present in the *Type5* (Fe5) sites, to avoid a higher migration energy. Hence, it is plausible to assume that the effect of the alloying elements with repulsive C-M binding energies on C diffusion can be highly concentration-dependent. Although the prior experimental results, Table 2, indicate that the alloying elements with repulsive C-M binding energies (Ni, Mn, Si, Al) enhance C migration in austenite (f.c.c.), our calculations, however, reveal that this is true only when the alloying elements occupy certain $1^{st}$ NN lattice sites of the C atom residing in an octahedral site, Figs. 9a-c, and if other certain $1^{st}$ NN lattice sites of the C atom, Figs. 9d-e, are occupied with these elements, the C atom must change the diffusion path due to the repulsive nature of C-M pairs, to avoid the higher energy states associated with the high-energy saddle-point. The comparison with the experiments indicates that the overall effects of these changes in the migration path on C diffusivity are perhaps negligible. Among the three austenite stabilizers (Ni, Mn, and Cu), Ni and Mn are C diffusion accelerators, while Cu is a carbon diffusion decelerator in the austenite (f.c.c.) phase. The vacillating elements, Cr and Mo, which can partition to austenite (f.c.c.) under certain conditions, are C diffusion decelerators in the f.c.c. phase.

**Table 2** The calculated migration energy of C in the austenite (f.c.c.) phase with and without the effects of alloying elements (Ni, Mo, V, Cr, Mn, Cu, Al, Ti, and Si), compared with the experimental data [13-16] and the calculated C-M binding energies (unit: eV).

| Alloying element (M): | -- | Ni | Mn | Ti | Si | Mo | Cr | Cu | Al | V |
|---|---|---|---|---|---|---|---|---|---|---|
| EXPT | 1.67 | 1.62 | 1.60 | - | 1.55 | 1.78 | 1.74 | - | 1.66 | - |
| DFT (M @ Fe0) | 1.82 | 1.64 | 1.57 | 2.17 | 1.51 | 1.86 | 1.94 | 1.91 | 1.74 | 2.07 |
| DFT (M @ Fe1) | 1.82 | 1.64 | 1.58 | 2.18 | 1.52 | 1.86 | 1.95 | 1.91 | 1.75 | 2.08 |
| DFT (M @ Fe2 or Fe3) | 1.82 | 1.68 | 1.59 | 2.19 | 1.53 | 1.88 | 1.97 | 1.92 | 1.77 | 2.09 |
| DFT (M @ Fe4) | 1.82 | 2.14 | 2.23 | 2.32 | 2.31 | 1.92 | 2.08 | 2.06 | 1.92 | 2.20 |
| DFT (M @ Fe5) | 1.82 | 2.10 | 2.21 | 2.22 | 2.25 | 1.85 | 2.00 | 1.97 | 1.89 | 2.11 |
| C-M Binding ($1^{st}$ NN) | -- | 0.22 | 0.27 | -0.28 | 0.45 | -0.08 | -0.16 | -0.14 | 0.09 | -0.24 |

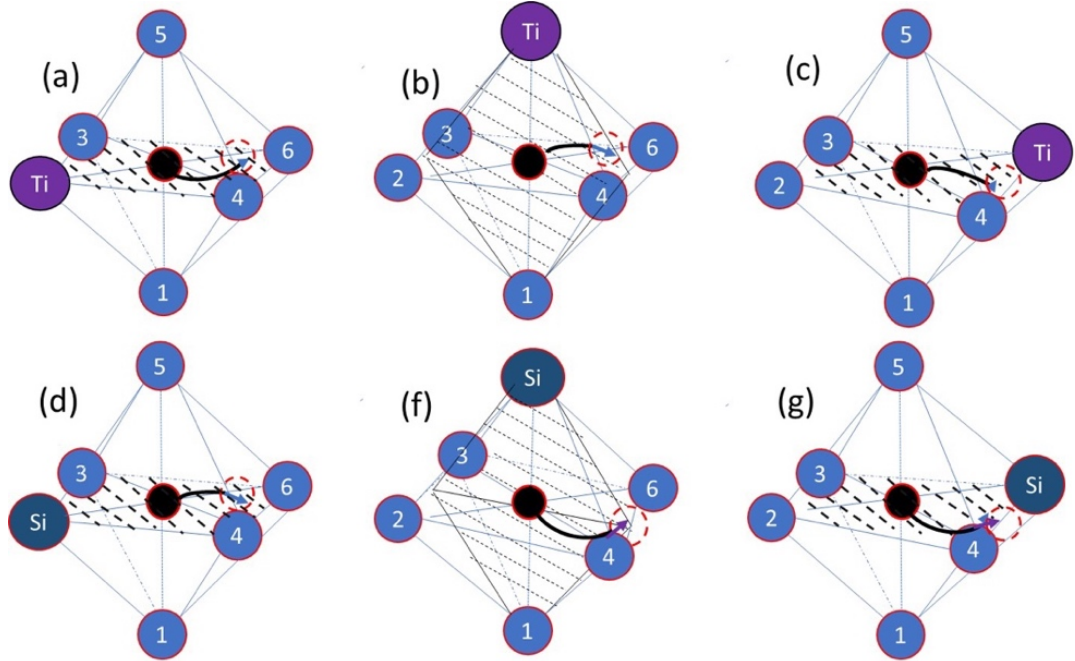

**Fig. 10** The C diffusion path under the influence of an alloying element at the effective 1$^{st}$ NN sites for both attractive (Ti) and repulsive (Si) binding-energies in ferrite (b.c.c.): (a) The attractive C-Ti binding energies at sites Fe2 or Fe3; (b) The attractive C-Ti binding energies at sites Fe1 or Fe5; (c) The attractive C-Ti binding energies at sites 4 or 6; (d) The repulsive C-Si binding energies at sites Fe2 or Fe3; (f) The repulsive C-Si binding energies at sites Fe1 or Fe5; and (g) The repulsive C-Si binding energies at sites Fe4 or Fe6. The plane of the C diffusion path is indicated by broken lines. Note the curvilinear paths as opposed to linear paths for the diffusing C atom. In (a), (c), (d), and (g) the {002} planes are cross-hatched (----), while in (b) and (f) the {020} planes are also cross-hatched (·····).

### 3.4.2. Carbon diffusion in b.c.c. ferrite under the effects of alloying elements

There are three different types of 1$^{st}$ NN sites for a C atom occupying an octahedral site in ferrite (b.c.c.), Fig. 6(a): the front-binding sites (Fe4 and Fe6), the in-plane-binding sites (Fe1 and Fe5), {002}, and the back-binding sites (Fe2 and Fe3). Our NEB calculations demonstrate that the presence of substitutional alloying elements at these different 1$^{st}$ NN sites alters the C migration path. With no alloying elements involved, C atoms diffuse along a vector from one distorted octahedral site to another NN octahedral site, through the first distorted tetrahedral transition site. This almost linear minimum-energy diffusion path in {002}-type planes changes to a curvilinear path in the {002}-type planes, due to the directionality of the C-M (attractive or repulsive) bonding. These changes in the C diffusion pathways are displayed in Fig. 10, for the case of Ti and Si solute atoms occupying the 1$^{st}$ NN sites of a C atom, which are determined employing the NEB methodology. Due to the curvilinear diffusion path, a C atom's migration path is as much as ~11% longer than a linear path. Below, we discuss the alloying elements with attractive and repulsive C-M binding energies separately.

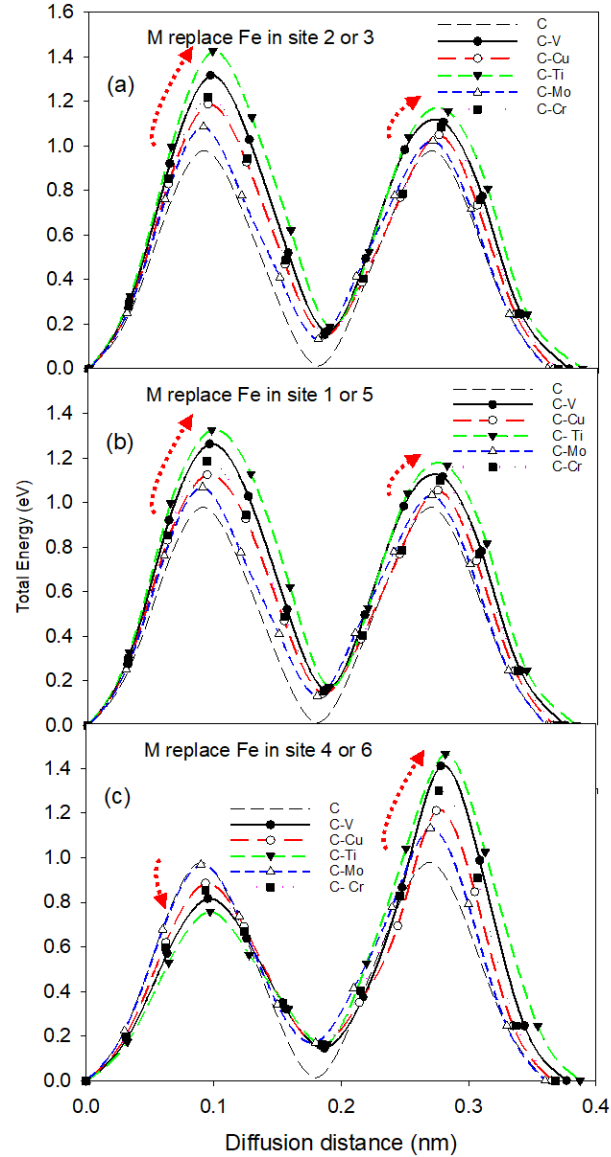

**Fig. 11** The effects of alloying elements with attractive C-M (V, Cu, Ti, Mo, Cr) binding energies on the migration energies of a carbon atom in ferrite (b.c.c.). (a) An alloying element substituting on sites Fe1 or Fe3; (b) An alloying element substituting on sites Fe1 or Fe5; and (c) An alloying element substituting on sites Fe4 or Fe6. The C-M binding energies introduce an asymmetry into the total energy versus diffusion distance plot. The red arrows indicate increasing or decreasing trends in the saddle-point energies.

*Effects of substitutional elements (V, Cu, Ti, Mo, Cr) with attractive C-M binding energies:* In the case of the back-binding sites, when an alloying element occupies Fe2 or Fe3, Fig. 6, we find that the 1$^{st}$ saddle-point energy increases significantly, but increases in the 2$^{nd}$ saddle-point energy are relatively small, Fig. 11a. Similar to the f.c.c. structure, the larger the attractive binding energy an alloying element has with the C atoms the greater is the corresponding saddle-point energy, Table 3. The C-Ti pair has the largest attractive binding energy at the 1$^{st}$ NN distance in ferrite (b.c.c.), -0.19 eV. In the presence of Ti at site Fe2 or site Fe3, the C migration energy increases from 0.98 eV in the Fe-C system to 1.42 eV for the 1$^{st}$ saddle-point and 1.15 eV for the 2$^{nd}$ saddle-point. The C-Mo pair, alternatively, has the smallest attractive binding energy at the 1$^{st}$ NN site, -0.08 eV, and C's migration energy is 1.08 eV for the 1$^{st}$ saddle-point and 1.02 eV for the 2$^{nd}$ saddle-point, when a Mo atom is present at site Fe2 or site Fe3.

In the case of in-plane-binding sites (Fe1 or Fe5), the migration energies are displayed in Fig. 11b. The C diffusion pathway is curvilinear within a {020}-type plane (Fe1 and Fe5) vertically cut through a {002} plane (Fe2, Fe3, Fe4 and Fe6) are displayed in Fig. 10b. It achieves a minimum energy when a C atom passes through the position between sites Fe4 and Fe6 with a migration distance of 0.22 to 0.28 nm. The largest migration energy, 1.33 eV, is for the 1$^{st}$ saddle-point and 1.17 eV for the 2$^{nd}$ saddle-point due to the effect of a Ti atom replacing Fe1 or Fe5. The migration energy is 1.06 eV for the 1$^{st}$ saddle-point and 1.03 eV for the 2$^{nd}$ saddle-point, due to the effect of a C-Mo pair, which has the smallest attractive binding energy.

In the case of the front-binding sites (Fe4 or Fe6), the alloying element is in front of C's diffusion path, The C diffusion-pathway is curvilinear with the alloying element atom nearby, which has opposite effects on the migration energies for the first and second saddle points. The 1$^{st}$ saddle-point decreases from its base value of 0.98 eV to 0.75 eV and the 2$^{nd}$ saddle-point energy increases from 0.98 eV to 1.47 eV, for a Ti atom occupying front-binding sites. These effects result in strongly trapping a C atom, which significantly decelerates C diffusion.

Overall, the C migration energies of C-M (V, Cu, Ti, Mo, Cr) increase from 0.98 eV for all six alloying sites, due to attractive C-M binding energies. We conclude that alloying elements with attractive C-M (V, Cu, Ti, Mo, Cr) binding energies decelerate C diffusion in ferrite (b.c.c.) caused by increasing migration energies for diffusion.

*Effects of substitutional elements (Ni, Mn, Si, and Al) with repulsive C-M binding energies:* We now consider the effects of alloying elements with repulsive C-M (Ni, Mn, Si, and Al) binding energies in ferrite (b.c.c.). Carbon's diffusion path is displayed in Fig. 12. The common feature of the three different types of substitutional sites is that C's diffusion path is curvilinear as opposed to linear. Sites Fe2 or Fe3, Fig. 12a, and site Fe1 or Fe5, Fig. 12b, have similar C diffusion paths. Site Fe4 or Fe6 is completely different from the other two types of sites. In the case of back-binding sites (Fe2 or Fe3), we find that the effects of alloying elements are significant on the 1$^{st}$ saddle-point and have less of an effect on the 2$^{nd}$ saddle-point when an alloying element occupies sites Fe2 or Fe3. The magnitude of decreases in saddle-point energies is proportional to the C-M binding energies. For instance, in the presence of Si (with a C-Si binding energy of 0.28 eV/pair), the first saddle-point energy decreases from 0.98 eV to 0.69 eV, and in the presence of Al (with a C-Al binding energy of 0.08eV/pair), it decreases slightly to 0.92 eV, Fig. 12a. The corresponding energies for the second saddle point are 0.87 and 0.97 eV, respectively.

In case of in-plane binding sites (Fe1 or Fe5), the migration energy trend is similar to that of the back-binding sites (Fe2 or Fe3). The C-M binding energies at sites Fe1 or Fe5 have significant effects on the 1$^{st}$ saddle-point, where the maximum migration energy decrease is from 0.98 eV to 0.75 eV for the C-Si binding energy: the in-plane diffusion path is curvilinear. The 2$^{nd}$ saddle-point energy decreases from 0.98 eV to 0.88 eV. The C-Al repulsive binding energy has the smallest value and concomitantly the migration energy is small, 0.93 eV, for the 1$^{st}$ saddle-point and 0.98 eV for the 2$^{nd}$ saddle -point.

**Table 3** The calculated migration energy of C in the ferrite (b.c.c) phase with and without the effects of alloying elements (Ni, Mo, V, Cr, Mn, Cu, Al, Ti, and Si), and the calculated C-M binding energies (unit: eV).

| Alloying element (M): | -- | Ni | Mn | Ti | Si | Mo | Cr | Cu | Al | V |
|---|---|---|---|---|---|---|---|---|---|---|
| EXPT | 0.87* | - | - | - | - | - | - | - | - | - |
| DFT (M @ Fe1 or Fe5) | 0.98 | 0.93 | 0.90 | 1,33 | 0.86 | 1.06 | 1.19 | 1.12 | 0.97 | 1.26 |
| DFT (M @ Fe2 or Fe3) | 0.98 | 0.91 | 0.88 | 1.44 | 0.84 | 1.08 | 1.21 | 1.18 | 0.96 | 1.32 |
| DFT (M @ Fe4 or Fe6) | 0.98 | 1.29 | 1.45 | 1.47 | 1.59 | 1.13 | 1.30 | 1.21 | 1.12 | 1.41 |
| C-M Binding (1$^{st}$ NN) | -- | 0.20 | 0.18 | -0.19 | 0.28 | -0.06 | -0.14 | -0.11 | 0.07 | -0.17 |

* C.A. Wert, Diffusion coefficient of C in α-iron, Phys. Rev. 79 (1950) 601–605.

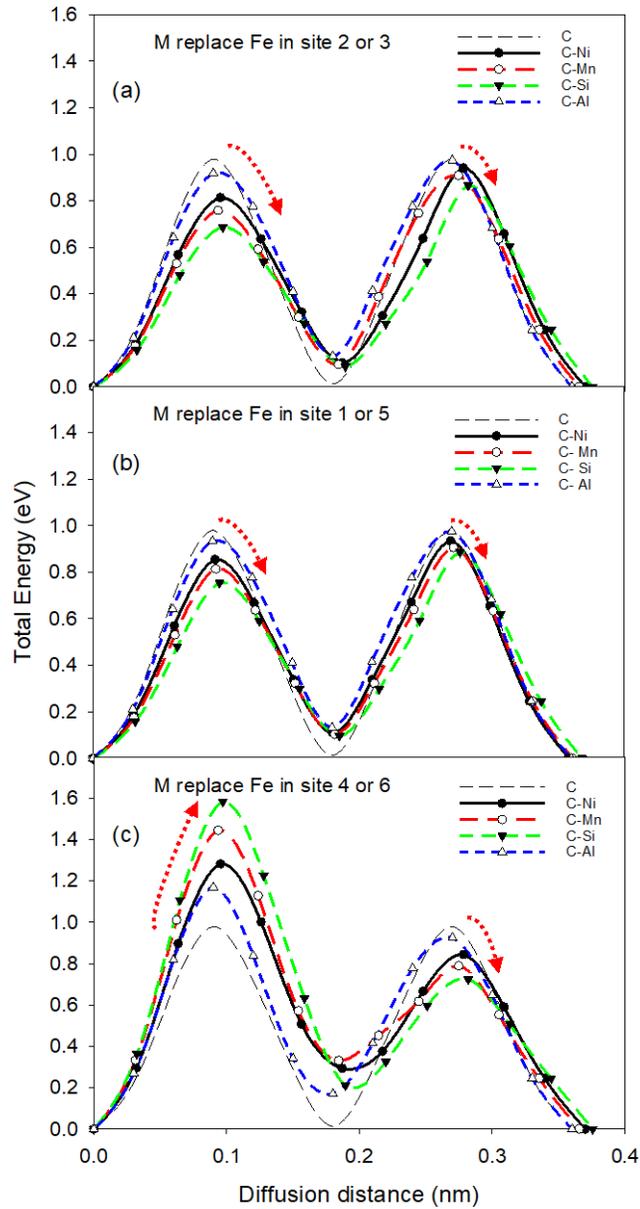

**Fig. 12** The effects of alloying elements with repulsive C-M (Ni, Mn, Si, Al) binding energies on the migration energies of a carbon atom in ferrite (b.c.c.). (a) An alloying element substituting on sites Fe1 or Fe3; (b) An alloying element substituting on sites Fe1 or Fe5; and (c) An alloying element substituting on sites Fe4 or Fe6: the C-M binding energies introduce an asymmetry into the total energy versus diffusion distance plot. The red arrows indicate increasing or decreasing trends in the saddle-point energies.

In case of the front-binding sites (Fe4 or Fe6), the alloying element blocks the C's diffusion-path. The repulsive C-M binding energy increases significantly the migration energy at the 1$^{st}$ saddle-point, Fig. 12c. When a C atom passes an alloying element, its migration energy decreases at the $2^{nd}$ saddle-point. The $1^{st}$ saddle-point increases from 0.98 eV to a maximum value of 1.58 eV for C-Si binding. The atomic distance between sites Fe4 and Fe6 increases by ~ 0.084 nm when C atoms migrate. The $2^{nd}$ saddle-point is 0.72 eV, which is smaller than the base value, 0.98 eV. We found that carbon atoms in ferrite (b.c.c.) have lower migration energy via in between Fe1-Fe5 or Fe2-Fe3, than Fe4-Fe6 as summarized in Table 3. We conclude that the repulsive C-M binding energy decreases the migration energy of C: the case of sites Fe2 or Fe3 or sites Fe1 or Fe5. These alloying elements accelerate C's diffusivity.

Overall, among the four ferrite stabilizers (Si, V, Al, Ti), Si is a C diffusivity accelerator in ferrite and V and Ti are C diffusivity decelerators in ferrite. Aluminum has a weakly repulsive binding energy with C and does not have a significant effect on C's diffusivity.

## 4. Summary and Conclusions

We studied the effects of alloying elements on diffusion pathways and migration energies of interstitial C atoms in austenite (f.c.c.) and ferrite (b.c.c.) using density functional theory (DFT) first-principles calculations and the nudged elastic band (NEB) method at 0 K: these calculations were made in the absence of vacancies. The elements studied are: Ni, Mo, V, Cr, Mn, Cu, Al, Ti, and Si, which are important alloying elements in 10 wt.% Ni steels. The following main conclusions are reached:

1. The binding conditions between C and the alloying elements are determined in the austenite (f.c.c.) and ferrite (b.c.c.) phases up to $6^{th}$ nearest-neighbor (NN) distances, Figs. 2-3: the 1st NN interactions are considered for migration energy calculations. The attractive or repulsive nature of the C-M pairs for all the elements studied is the same in both austenite (f.c.c.) and ferrite (b.c.c.) phases, but the absolute values of the binding energies in austenite (f.c.c.) are much larger than those in ferrite (b.c.c.). Nickel, Mn, Al, and Si have positive repulsive binding energies, while Mo, V, Cr, Cu, and Ti have negative attractive binding energies in both austenite (f.c.c.), Fig. 2, and ferrite (b.c.c.), Fig. 3. Carbon-Si has the largest positive binding energy: 0.28 eV in ferrite and 0.45 eV in austenite. Carbon-Ti has the largest negative binding energy: -0.19 eV in ferrite and -0.28 eV in austenite.

2. Carbon diffusion paths and migration energies are studied in detail. In austenite, Fig. 4, C diffusion is from an initial octahedral interstitial site to another NN octahedral site through a transition tetrahedral site and two intermediate saddle-point sites. And the calculated C migration energy is 1.82 eV, Fig. 5. The diffusion path is through a tetrahedral interstitial site, a curvilinear path connecting two octahedral sites. In ferrite, Fig. 6, C diffusion is from an octahedral interstitial site to another NN octahedral site through a distorted tetrahedral transition site, a distorted octahedral site, and another distorted tetrahedral transition site. There are two saddle points at the distorted tetrahedral transition site. The calculated migration energy is 0.98 eV, Fig. 7. The C diffusion path is linear and connects the two regular octahedral interstitial sites.

3. The alloying elements in Fe have effects on both the C diffusion pathway and the migration energies. The elements with attractive binding energies decelerate C diffusion, while the elements with repulsive binding energies accelerate C diffusion. In ferrite (b.c.c.), the C diffusion path is curvilinear in the diffusion plane instead of a linear path under the effects of alloying elements.

4. Among the three austenite stabilizers studied (Ni, Mn, and Cu), Ni and Mn are C diffusion accelerators in austenite, while Cu decelerates C's diffusivity in austenite.

5. Our APT analyses, Fig. 1, reveals that the vacillating elements, Cr and Mo, also partition to austenite in a QLT-treated 10 wt.% Ni steel and are C diffusion decelerators in austenite.

6. Among the four ferrite stabilizers (Si, V, Al, Ti), Si is the C diffusion accelerator in ferrite, and V and Ti are C diffusion decelerators in ferrite. Aluminum has a weak repulsive binding energy with carbon and does not have a significant effect on C's diffusivity.

**Acknowledgments**

We are grateful to Dr. Divya Jain and Prof. Dieter Isheim for numerous valuable discussions on atom-probe tomography analyses of steels. This research was supported by the Office of Naval Research (N000141812594), grant officer Dr. W. Mullins. Atom-probe tomography was performed at the Northwestern University Center for Atom-Probe Tomography (NUCAPT). The LEAP tomograph at NUCAPT was purchased and upgraded with grants from the NSF-MRI (DMR-0420532) and ONR-DURIP (N00014-0400798, N00014-0610539, N00014-0910781, N00014-1712870) programs. NUCAPT received support from the MRSEC program (NSF DMR-1720139) at the Materials Research Center, the SHyNE Resource (NSF ECCS-1542205), and the Initiative for Sustainability and Energy (ISEN) at Northwestern University. This research was supported in part through the computational resources and staff contributions provided for the Quest high-performance computing facility at Northwestern University, which is jointly supported by the Office of the Provost, the Office for Research, and Northwestern University Information Technology.